\documentclass{IEEEtran}
\usepackage{cite}
\usepackage{amsmath,amssymb,amsfonts}
\usepackage{algorithmic}
\usepackage{graphicx}
\usepackage{epstopdf}
\usepackage{textcomp}
\usepackage{multirow}
\usepackage{booktabs}
\usepackage{amsmath}
\usepackage{cite}
\usepackage{color}

\def\BibTeX{{\rm B\kern-.05em{\sc i\kern-.025em b}\kern-.08em
    T\kern-.1667em\lower.7ex\hbox{E}\kern-.125emX}}
\begin{document}
\title{A Surface Integral Formulation for Scattering Modeling by 2D Penetrable Objects}
\author{Xiaochao Zhou, \IEEEmembership{Graduate Student Member, IEEE}, Shunchuan Yang, \IEEEmembership{Member, IEEE},\\ and Donglin Su, \IEEEmembership{Member, IEEE}
\thanks{Manuscript received xxx; revised xxx.}
\thanks{This work was supported in part by the National Natural Science Foundation of China through Grant 61801010, Grant 61727802, the ``$13$th Five-Year" Equipment Pre-research Fund through Grant 61402090602 and Beijing Natural Science Foundation through Grant 4194082. \textit{(Corresponding author: Shunchuan Yang})}
\thanks{All authors are with the School of Electronic and Information Engineering, Beihang University, Beijing, 10083, China (e-mail: zhouxiaochao@buaa.edu.cn, scyang@buaa.edu.cn, sdl@buaa.edu.cn).}
}

\maketitle

\begin{abstract}
In this paper, we proposed a single-source surface integral formulation to accurately solve the scattering problems by 2D penetrable objects. In this method, the objects are replaced by their surrounding medium through enforcing a surface equivalent electric current to ensure fields exactly the same as those in the original scattering problem.  The equivalent electric current  is obtained through emitting the equivalent magnetic current by enforcing that the electric fields in the original and equivalent problems equal to each other. Through solving the Helmholtz equation inside objects by the scalar second Green theorem, we could accurately model arbitrarily shaped objects. Then, we solve the exterior scattering problems through the combined integral equation (CFIE) with the equivalent electric current. The proposed formulation only requires a single electric current source to model penetrable objects. At last, two numerical experiments are carried out to validate its accuracy, stability and capability of handling non-smoothing objects.
\end{abstract}


\section{Introduction}
\label{sec:introduction}
\IEEEPARstart{T}{he} method of moment (MOM), which is based on surface integral equations, is widely used to solve scattering problems \cite{MLFMP} and extract the electrical parameters of large-scale integrated circuits (ICs) \cite{FastImp}. Compared with other volumetric methods, like the finite-difference time-domain (FDTD) method \cite{FDTD} and the finite element method (FEM) \cite{FEM}, it has much smaller count of unknowns. 

To model the piecewise homogenous penetrable media, various formulations, like the Poggio-Miller-Chan-Harrington-Wu-Tsai (PMCHWT) \cite{PMCHWT}, the combined tangential field (CTF) \cite{CTF}, are proposed in the last decades. In those formulations, a two-region problem is solved by introducing both the surface equivalent electric and magnetic currents through the equivalence theorem. Therefore, it would be desirable to achieve single-source formulations to model penetrable objects. When considering the conducting media, the single-source formulation could be obtained through incorporating with the impedance boundary condition \cite{IMBC} or the generalized impedance boundary condition \cite{GIBC}. Further investigations found that they suffer from accuracy issue in the low frequency range or inner resonance problem. Recently, new surface-volume-surface single-source formulations through mapping volume integral operator to its surface counterpart were proposed to model penetrable objects \cite{SVS, SVS2}. They could significantly improve the efficiency compared with the volume integral equation methods. However, they still involve the volume integral operator.

In this paper, we proposed another single-source formulation to solve the 2D transverse magnetic (TM) scattering problems for penetrable objects. The proposed formulation is based on the combined integral equation (CFIE) incorporated with the equivalent surface electric current obtained from the differential surface admittance operator, which was introduced in \cite{SOPZUTTER} to model high-speed interconnects. Many other efforts are made to extend its capabilities such as modeling circular cables \cite{SOPCIRCULAR}, arbitrarily shaped interconnects \cite{SOPCIM}, 3D scattering problems \cite{SCATTERING} and antenna array \cite{ARRAYS}. In this paper, we further extend the differential surface admittance operator to solve the  2D TM scattering problems by arbitrarily shaped penetrable objects. In the proposed formulation, only the surface equivalent electric current source is required. 

The paper is organized as follows. In Section II, we demonstrate the proposed formulation with the single equivalent electric current source in detail. In Section III, numerical experiments are carried out to validate its accuracy, stability and capability of handling non-smoothing objects. At last, we draw some conclusions.

\section{Formulation}
\subsection{The Equivalent Surface Current}
We consider a penetrable object denoted its boundary as $\gamma$ and interior domain as $S$, with the permittivity, the permeability and the conductivity as ${\varepsilon _1}$, ${\mu _1}$,  ${\sigma_1}$, respectively. It is surrounding by a background medium with ${\varepsilon _0}$, ${\mu _0}$,  ${\sigma_0}$ and illuminated by a plane wave as shown in Fig. 1(a). According to the surface equivalence theorem, we could obtain an equivalent problem, in which fields are exactly the same as those in the original problem through replacing the objects by their surrounding medium and enforcing a surface equivalent electric and magnetic current at $\gamma$ as shown in Fig. 1(b) \cite{SOPZUTTER}. The exterior fields in the original problem are exactly the same as those in the equivalent problem, denoted as ${E_0=\widehat{ E}_0}$ and ${H_0=\widehat {H}_0}$. These two surface currents are nothing but the difference between the tangential fields in the original and equivalent problems. If we carefully select the electric and magnetic fields in the equivalent problem, a single source formulation could be obtained.  As stated in \cite{SOPZUTTER}, a single source formulation by carefully choosing the electric fields in the equivalent problem is achieved. In this paper, we consider the single surface equivalent electric current, denoted as $\widehat{\bf{J}}_s$. However, we could get its dual single source formulation for the magnetic source using the same manner. 

\begin{figure}
	\begin{minipage}[h]{0.48\linewidth}\label{FIG2A}
		\centerline{\includegraphics[width=1.65in]{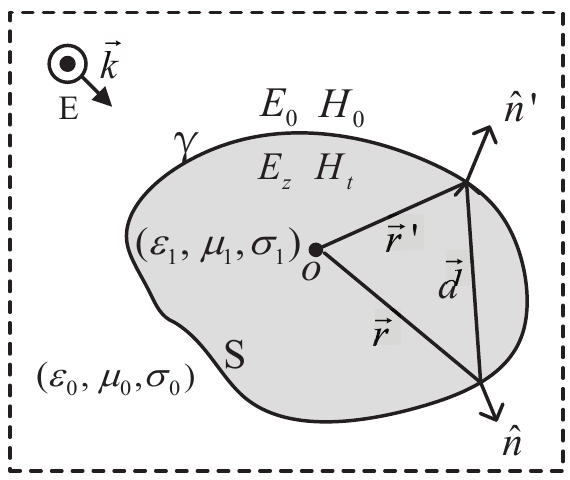}}
		\centerline{(a)}
	\end{minipage}
	\hfill
	\begin{minipage}[h]{0.48\linewidth}\label{FIG2B}
		\centerline{\includegraphics[width=1.65in]{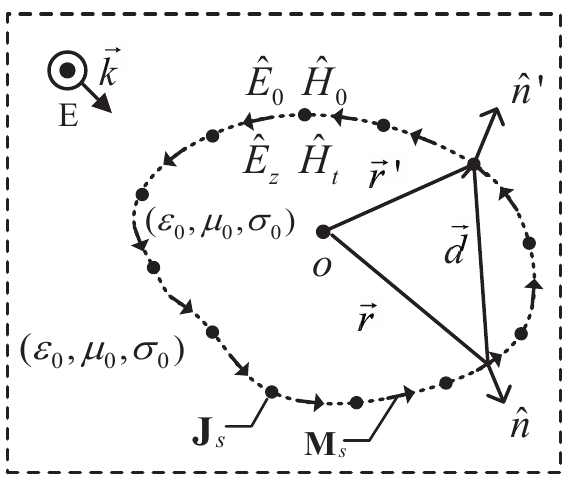}}
		\centerline{(b)}
	\end{minipage}
	\caption{(a) Original scattering problem, where $E_0$ and $H_0$ are exterior fields and $E_z$, $H_t$ are interior fields and (b) the equivalent problem with the object replacing by its surrounding medium and enforcing the surface electric and magnetic currents at $\gamma$, where $\widehat {E}_0$, $\widehat {H}_0$, $\widehat {E}_z$ and $\widehat {H}_t$ are exterior and interior fields, respectively.}
	\label{fig_2}
\end{figure}

According to the equivalence theorem \cite{EQUIBOOK}, the surface equivalent electric current and magnetic current could be expressed as 
\begin{equation}{\label{EQUEM}}
	{\widehat{ \bf{J}}_s(\vec r')} = {\bf{H}}_t(\vec r') - \widehat {\bf{H}}_t(\vec r'),  {\widehat{\bf{M}}_s(\vec r')} = {\bf{E}}_t(\vec r') - \widehat {\bf{E}}_t(\vec r')
\end{equation}
where ${\bf{E}}_t(\vec r')$, $\widehat {\bf{E}}_t(\vec r')$, ${\bf{H}}_t(\vec r')$, $\widehat {\bf{H}}_t(\vec r')$ are the surface tangential electric and magnetic fields in the original and equivalent problems, respectively and all quantities with a hat denote their values in the equivalent problem.

Since the electric and magnetic fields in the equivalent problem could be arbitrary, we could obtain the single electric current source by enforcing that ${\bf{E}}_t(\vec r') = \widehat {\bf{E}}_t(\vec r')$ and have 
\begin{equation}{\label{SINGLEJ}}
\widehat {\bf{J}}_s(\vec r') \neq {\bf{0}},  \widehat {\bf{M}}_s(\vec r') = {\bf{0}}. 
\end{equation}
On the other hand, we could have the single magnetic current source by enforcing that ${\bf{H}}_t(\vec r') = \widehat {\bf{H}}_t(\vec r') $ and have 
\begin{equation}{\label{SINGLEM}}
\widehat {\bf{M}}_s(\vec r') \neq {\bf{0}},\widehat {\bf{J}}_s(\vec r') = {\bf{0}}. 
\end{equation}
Both of them could give us a single-source formulation. In this paper, we select the first to obtain the single-source integral formulation and then derive the surface equivalent electric current using the contour integral method \cite{CIPBOOK}. 

Considering the penetrable objects do not include any sources, the electric fields must satisfy the following scalar Helmholtz equation in $S$,
\begin{equation}{\label{HELMHOLTZ}}
	{\nabla ^2}E_z + {k^2}E_z = 0,
\end{equation}
subject to the boundary condition,
\begin{equation}{\label{BOUNDARYC}}
	E_z(\vec r) = \widehat {E}_z(\vec r), \quad \vec r \in {\gamma }.
\end{equation}
({\ref{HELMHOLTZ}}) could be solved through the second scalar Green theorem, and we obtain the relationship between the electric field $E_z$ and its normal derivative of $E_z$ at $\gamma$ \cite{CIPBOOK},
\begin{equation}{\label{INTEGRALEQU}}
	\frac{1}{2} E_z(\vec r) = \oint_{{\gamma }} {\left[ G(\vec r,\vec r')\frac{{\partial E_z(\vec r')}}{{\partial n'}}- {\frac{{\partial G(\vec r,\vec r')}}{{\partial n'}}E_z(\vec r')} \right]} dr',
\end{equation}
where $G$ is the Green function expressed as $G = -\frac{i}{4} H_0^{(2)}(k\rho)$, where $k$ is the wavenumber in the object and $H_0^{(2)}(\cdot)$ is the zeroth-order Hankel function of the second kind. In addition, the tangential magnetic field relates to the electric field in 2D-TM at $\gamma$ through the Poincare-Steklov operator \cite{SOPZUTTER} as
\begin{equation}{\label{HTM}}
	H_t(\vec r) =  {\frac{1}{{j\omega \mu }}{{\left[ {\frac{{\partial E _z(\vec r)}}{{\partial n}}} \right]}_{\vec r \in {\gamma}}}},
\end{equation}
where $\mu$ is the permeability of the object. We use the pulse basis function to expand $E_z$ and $H_t$ and point-matching scheme to test (\ref{INTEGRALEQU}) and (\ref{HTM}) at midpoints of each segment of $\gamma$. Then, after substituting (\ref{HTM}) into (\ref{INTEGRALEQU}), we collect all $E_z$ and $H_t$ coefficients into two column vectors,
$\bf{E}$, $\bf{H}$, and write (\ref{INTEGRALEQU}) into matrix form as
\begin{equation}{\label{PU}}
	{\bf{PH}} = {\bf{UE}}. 
\end{equation}
Next, through inversing the square matrix ${\bf{P}}$, we obtain the surface admittance operator $\bf{Y}$ as
\begin{equation}{\label{YY}}
	{\bf{H}} = \underbrace {{\bf{P}}^{{\bf{ - 1}}}{{\bf{U}}}}_{\bf{Y}}{\bf{E}}.
\end{equation}
$\bf{E}$, $\bf{H}$ are denoted as the discretized $E_z$, $H_t$ at the interior of $\gamma$ in the original problem. 
When all the parameters are replaced by its surrounding medium, we obtain the equivalent problem. With similar procedure, for the equivalent problem, we could obtain 
\begin{equation}
	{\widehat{\bf{H}}} = \underbrace {{\widehat{ \bf{P}}}^{ - 1}{ \widehat{\bf{U}}}}_{\widehat{ \bf{Y}}}\bf{E},
\end{equation}
where $\widehat{\bf{H}}$ represents the discretized magnetic field ${{\widehat H}_t}$ in the equivalent problem, as shown in Fig. 1(b).   Since we have enforcing (\ref{SINGLEJ}), we have 
\begin{equation}{\label{YSDEF}}
	{{\bf{\widehat{J}}}_s} = \bf{{Y_s}}\bf{{E}},
\end{equation}
where $\bf{Y_s}$ is the differential surface admittance operator and could be expressed as
\begin{equation}{\label{YS}}
	{\bf{Y_s}} =\bf{Y} - \widehat{\bf{Y}}  ={\bf{P}^{ - 1}}\bf{U} - {\widehat{ \bf{P}}}^{ - 1}{ \widehat{\bf{U}}}.
\end{equation}

When multiple scatters are involved, the $\bf{Y_s}$ is a block diagonal matrix assembling from all differential surface admittance operator for each object.

\subsection{The Scattering Modeling}
The scattering fields induced by ${\bf{J}}_s$ at $\gamma$ in the exterior region of the equivalent problem \cite{MOMBOOK} could be expressed as
\begin{align}
&{\bf{E}^s}(\vec r) =  - j\omega {\mu_0} {\int_{\gamma } {{\bf{J}}_s(\vec r'){G_0}(\vec r,\vec r')ds'} }, {\label{EFIE}}\\
&{\bf{H}^{s}}( \vec r ) = -{\bf{\hat n}} \times \frac{{{{\bf{J}}_s}}}{2} - \int_{\gamma } {{{\bf{J}}_s} \times \nabla {G_0}\left( {{{\vec r}_0},\vec r'} \right)d\vec s'}, {\label{MFIE}}
\end{align}
where $\vec r' \in {\gamma}$, $\omega$, $\mu_0$ and $\bf{\hat{n}}$ denote the angular frequency, the permittivity of the background medium, and the unit normal vector pointing out of the object, respectively, ${G_0}(\vec r,\vec r')$ is the Green function expressed as ${G_0}(\vec r,\vec r') = -\frac{j}{4}H_0^{(2)}(k_0 \rho)$, where $k_0$ is the wavenumber in the background medium. Unlike for the interconnect problems, where a 2D static Green function is used with the quasi-static assumption, the full wave Green function should be used for scattering modeling.

Therefore, the total fields at the exterior region could be expressed as the superposition of the scattering and incident fields, ${{\bf{E}}_0 = {{\bf{E}}_s} + {{\bf{E}}^{inc}}}$ and ${{\bf{H}}_0 = {{\bf{H}}_s} + {{\bf{H}}^{inc}}}$. We could rewrite them as 
\begin{align}
&{{\bf{E}}_{o}} =  {\widehat{\textit{L}}}({{\bf{J}}_s}) + {{\bf{E}}^{inc}} {\label{EFIEMAT}}, \\
&{{\bf{H}}_{o}} =  {\widehat{\textit{K}}}({{\bf{J}}_s}) + {{\bf{H}}^{inc}} {\label{MFIEMAT}},
\end{align}
where 
\begin{align}
&\widehat{L}({{\bf{J}}_s}) =  - j\omega {\mu_0} {\int_{{\gamma }} {{{\bf{J}}_s}(\vec r'){G_0}(\vec r,\vec r')ds'} }, \\
&\widehat{K}({{\bf{J}}_s} ) = -{\bf{\hat n}} \times \frac{{{{\bf{J}}_s}}}{2} - \int_{\gamma} {{{{\bf{J}}_s}} \times \nabla {G_0}\left( {{{\vec r}_0},\vec r'} \right)d\vec s'}. 
\end{align}

To avoid possible internal resonance, we select to use the linear combination of (\ref{EFIEMAT}) and (\ref{MFIEMAT}). Therefore, we have the following formulation, 
\begin{equation}{\label{KOP}}
\begin{aligned}
\alpha {{\bf{E}}_{o}} + \left( {1 - \alpha } \right){\bf{\hat n}} \times \eta {{\bf{H}}_{o}} = \alpha \left( {{{\bf{E}}^s} + {{\bf{E}}^{inc}}} \right)  \\
+ \left( {1 - \alpha } \right){\bf{\hat n}} \times \eta \left( {{{\bf{H}}^s} + {{\bf{H}}^{inc}}} \right),
\end{aligned}
\end{equation}
where $\alpha$ is a real constant in $[0,1]$ and $\eta$ denotes the wave impedance of the background medium. $\alpha$ is selected as 0.5 in all simulations.

We expand ${\bf{E}}_{o}$ and ${\bf{H}}_{o}$ with the pulse basis function and use the point-matching scheme at each segment of $\gamma$. Since the surface electric field at $\gamma$ in the equivalent problem has been enforced to equal to that in the original problem, $\bf{Y_s}$ relates $\bf{J_s}$ and $\bf{E}$ through (\ref{YS}), where $\bf{Y_s}$ is introduced in \cite{SOPZUTTER,SOPCIM} for interconnect modeling. It should be noted that \cite{SOPSCATTERDTWO} explore the possibility to use the surface admittance operator to solve the scattering problems. However, the method in \cite{SOPSCATTERDTWO} is only applicable for the canonical objects and possibly suffers from internal resonance issues. In this paper, the proposed method is free from internal resonance and applicable for arbitrarily shaped objects. 

 With the definition of $\bf{Y}$ in (\ref{YY}) and $\bf{Y_s}$ in (\ref{YS}), we could solve the electric field $\bf{E}$ at ${\gamma}$ through (\ref{KOP}) as
\begin{equation}{\label{CFIEFINAL}}
\begin{aligned}
{\bf{E}}  =&  \left[ {\alpha \left( {{\bf{I}} + {\bf{\widehat{L}Y_s}}} \right) + \left( {1 - \alpha } \right)\eta \left( {{\bf{Y}} + {\widehat{\bf{K}}}{{\bf{Y}}_s}} \right)} \right]^{ - 1} \\
		          & \left[ {\alpha {{\bf{E}}^{inc}} + \left( {1 - \alpha } \right){\bf{\hat n}} \times \eta {{\bf{H}}^{inc}}} \right],
\end{aligned}
\end{equation}
where $\bf{I}$ is identity matrix.  Readers should be careful that $\bf{Y}$ in (\ref{CFIEFINAL}) rather than $\widehat{\bf{Y}}$ should be used. Since no currents at $\gamma$ exist, the magnetic fields at inner and exterior side of $\gamma$ in the original problem equal to each other as shown in Fig. 1(a), i.e ${H_t}(\vec r') = {H_0}(\vec r')$, $\vec r' \in \gamma $. In the equivalent problem, we have ${\widehat H_0}(\vec r') = H_0(\vec r')$. Therefore, Y should be used.

\subsection{Electric Field Computation}
There are three domains, the exterior region of objects, the boundary and inner region of objects in the computational domain. When we require the electric fields in the exterior region of objects, they could be obtained through (\ref{EFIE}) and (\ref{YSDEF}). At $\gamma$, we already have the fields at the middle points of each segment. Therefore, the fields at other locations of $\gamma$ could be easily interpolated through calculated values. In $S$, we obtain fictions fields since we use the equivalence theorem. We could  use the tangential electric fields at $\gamma$ to calculate the fields in the interior region of objects as,
\begin{equation}
{{\bf{E}}_{in}} = \left({\bf{U'}} - {\bf{{P'}Y}} \right) \bf{E},
\end{equation}
where $\bf{Y}$ is exactly the same as that in (\ref{YY}), $\bf{P'}$, $\bf{U'}$ are obtained from the first and the second term of right hand side of (\ref{INTEGRALEQU}), respectively.

\section{Numerical Results and Discussion}
In this section, we present two numerical examples to demonstrate the accuracy, stability and the capability of handling non-smoothing objects.

\begin{figure}
	\begin{minipage}[h]{0.48\linewidth}\label{FIG2A}
		\centerline{\includegraphics[width=2.0in]{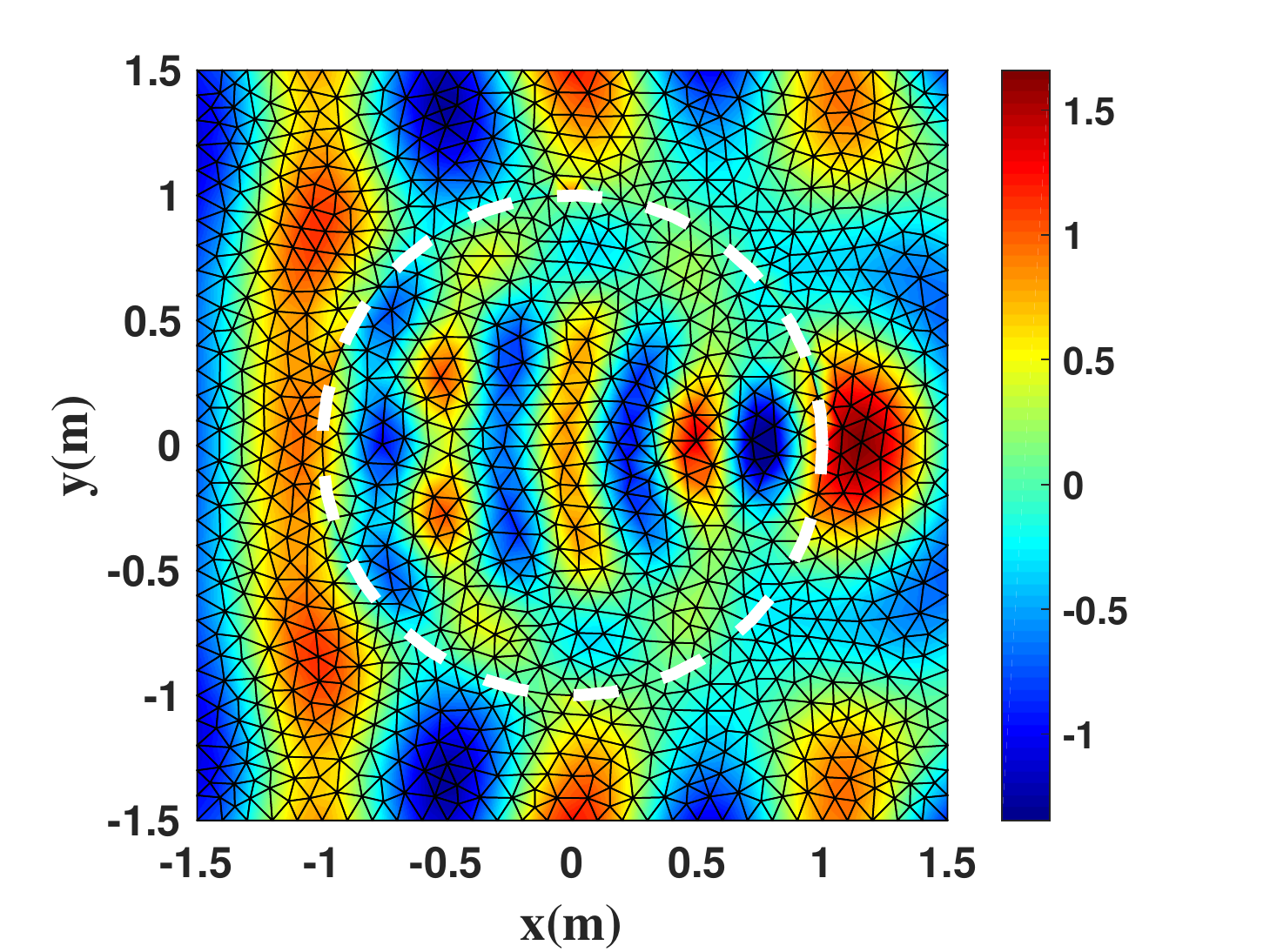}}
		\centerline{(a)}
	\end{minipage}
	\hfill
	\begin{minipage}[h]{0.48\linewidth}\label{FIG2B}
		\centerline{\includegraphics[width=2.0in]{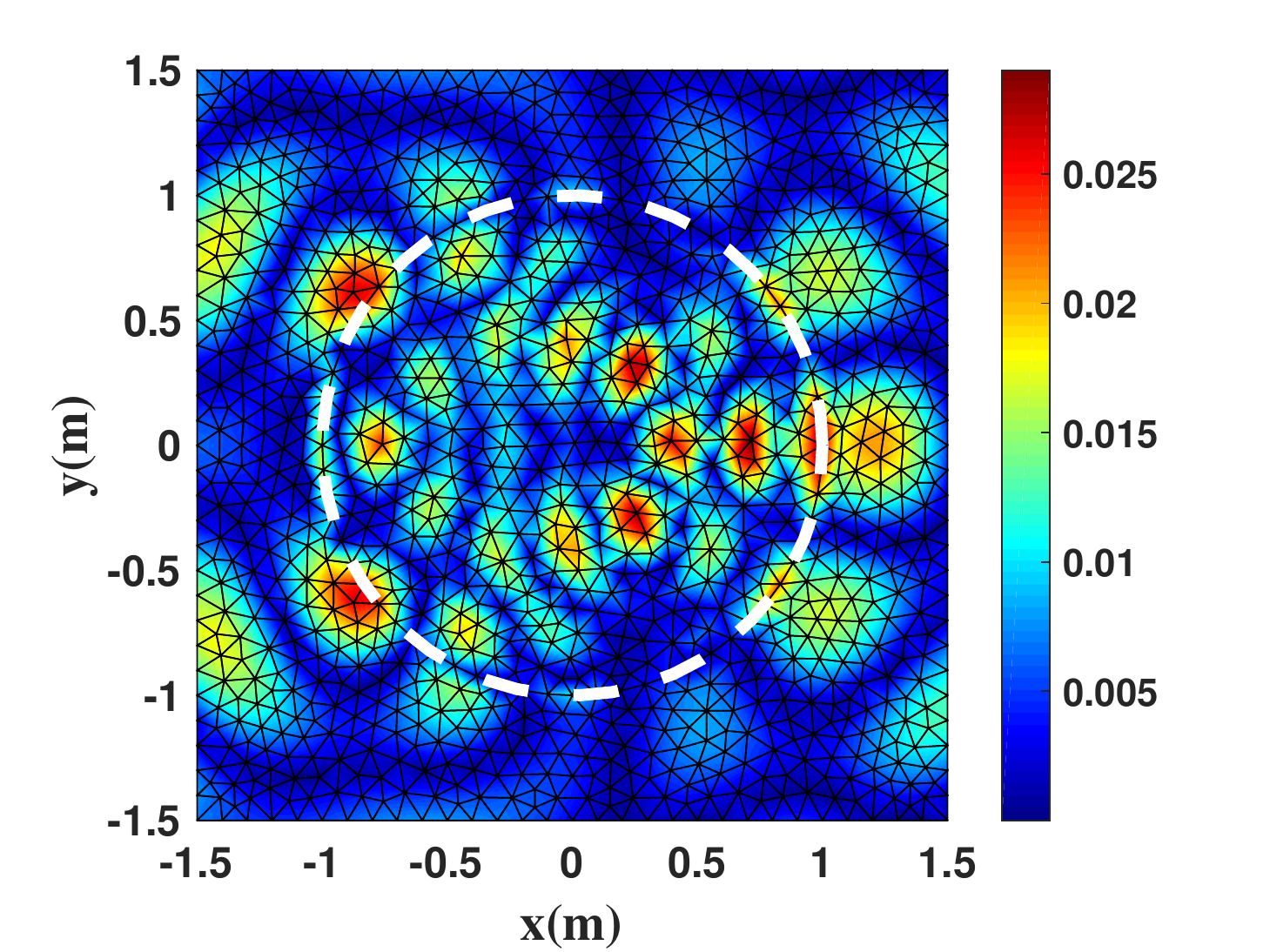}}
		\centerline{(b)}
	\end{minipage}
	\begin{minipage}[h]{0.48\linewidth}\label{FIG2A}
		\centerline{\includegraphics[width=1.8in]{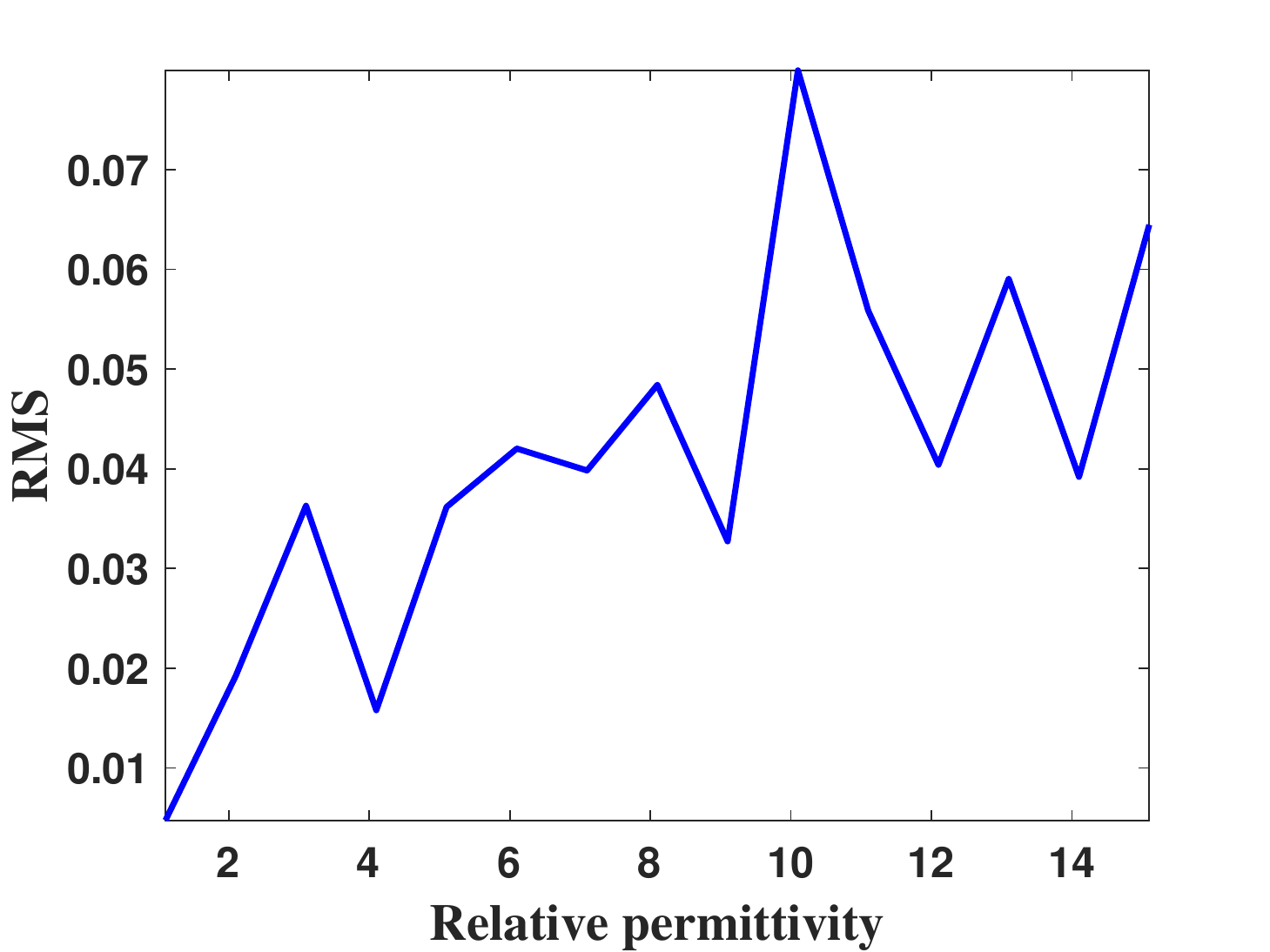}}
		\centerline{(c)}
	\end{minipage}
	\hfill
	\begin{minipage}[h]{0.48\linewidth}\label{FIG2B}
		\centerline{\includegraphics[width=1.8in]{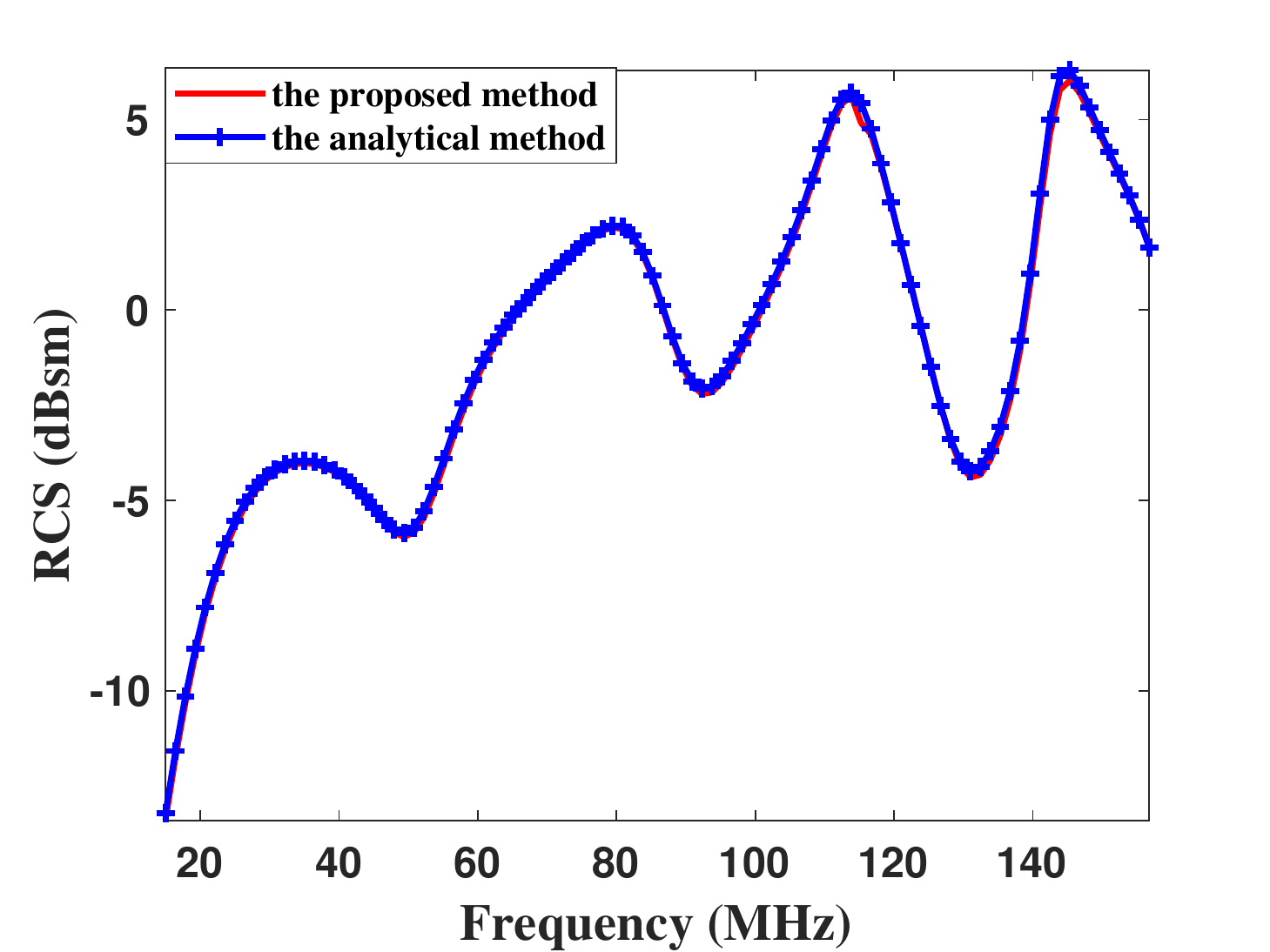}}
		\centerline{(d)}
	\end{minipage}
	\caption{(a) Electric fields obtained from the proposed method, (b) relative error between electric field obtained from the proposed method and the analytical method, (c) relative error verse the permittivity of the object and (d) relative error of the RCS obtained from the proposed method and the analytical method in a wide frequency range.}
	\label{fig_2}
\end{figure}

\subsection{A Dielectric Cylinder Object}
The first numerical experiment considered is a infinite cylinder with $\epsilon_r = 4$. Its radius is one wavelength, $1\lambda_0$, where $\lambda_0$ is the wavelength in the free space. The background medium is air. A plane wave with the frequency 300 MHz incidents from the $x$ axis. The minimum segment length is selected as $\lambda_0 / 10$, 0.1 m to discretize the contour of the cylinder. In the following simulations, we will use this configuration otherwise stated. The relative error is defined as $|E^{ref}-E^{cal}|/max|E^{ref}|$, where $E^{ref}$ and $E^{cal}$ denote the reference and calculated fields. The analytical fields are selected as the reference.

We obtain the electric fields in the computational domain as shown in Fig. 2 (a) and the error pattern is shown as in Fig. 2(b). It is easy to find that the maximum relative error is below 3\% compared with the analytical fields. It shows that the proposed method could obtain accurate results for the penetrable objects. The error could be further reduced as we refine the mesh.  

We further investigate the accuracy and stability of the proposed method for different $\epsilon_r$. For convenience, we fixed the mesh size to 0.05 m for the wideband analysis. The error is defined as RMS$=\sqrt {{{\sum {{{\left| {{E^{ref}} - {E^{cal}}} \right|}^2}} }}/{{\sum {{{\left| {{E^{ref}}} \right|}^2}} }}}$. As shown in Fig. 2(c), we  find that the proposed method could get accurate results with $\epsilon_r$ changing from 1.1 to 15. In addition, as expected, the error will rise as the permittivity of the object increases since in our simulations we fixed the mesh size. For the medium with a high permittivity, the mesh could not be enough sampled. However, the error could be further reduced as the mesh is refined. Therefore, the proposed method is accurate and stable from the low to high $\epsilon_r$ values. The results show that the proposed method has excellent performance in terms of accuracy and stability for different media parameter contrast.

Fig. 2(d) illustrates the accuracy of the proposed method in a wide frequency range. We fixed the mesh size at 0.05 m. The RCS obtained from the proposed method and the analytical solution shows good agreement from 15 MHz to 150 MHz. Therefore, the proposed method could obtain accurate results in a wideband frequency range.

\subsection{A Dielectric Cuboid Object}
An infinite long cuboid with non-smoothing corners and $\epsilon_r = 4$ is considered. Its surrounding medium is air. The incident plane wave is along the $x$ axis with the frequency 300 MHz. The side length is $1 \lambda_0$, 1 m. The minimum mesh size is 0.05 m to discretize the contour of the cuboid.

Fig. 3(a) shows the relative error between the electric fields obtained from the proposed method and the Comsol. It is easy to find that the relative error in the whole computational domain is less than 3\%. It implies that the proposed method could get accurate fields compared with the Comsol. Meanwhile, the RCS obtained through both methods shows good agreement as shown in Fig. 3(b). 

As shown in these numerical experiments, the proposed method is applicable objects with not only smoothing surfaces but also non-smoothing surfaces and shows good performance in terms of accuracy and stability.

\begin{figure}
	\begin{minipage}[h]{0.48\linewidth}\label{FIG2A}
		\centerline{\includegraphics[width=1.95in]{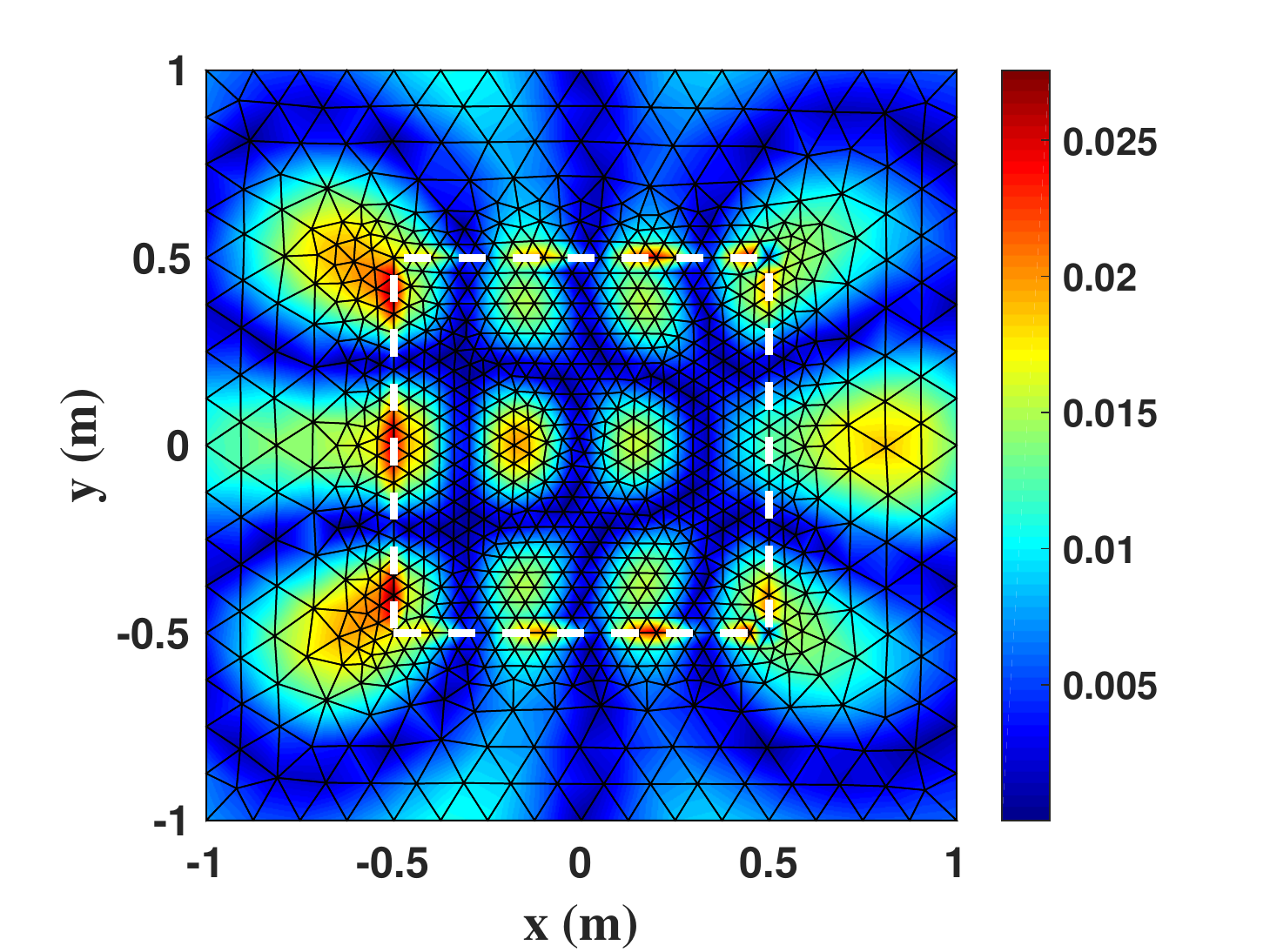}}
		\centerline{(a)}
	\end{minipage}
	\hfill
	\begin{minipage}[h]{0.48\linewidth}\label{FIG2B}
		\centerline{\includegraphics[width=1.85in]{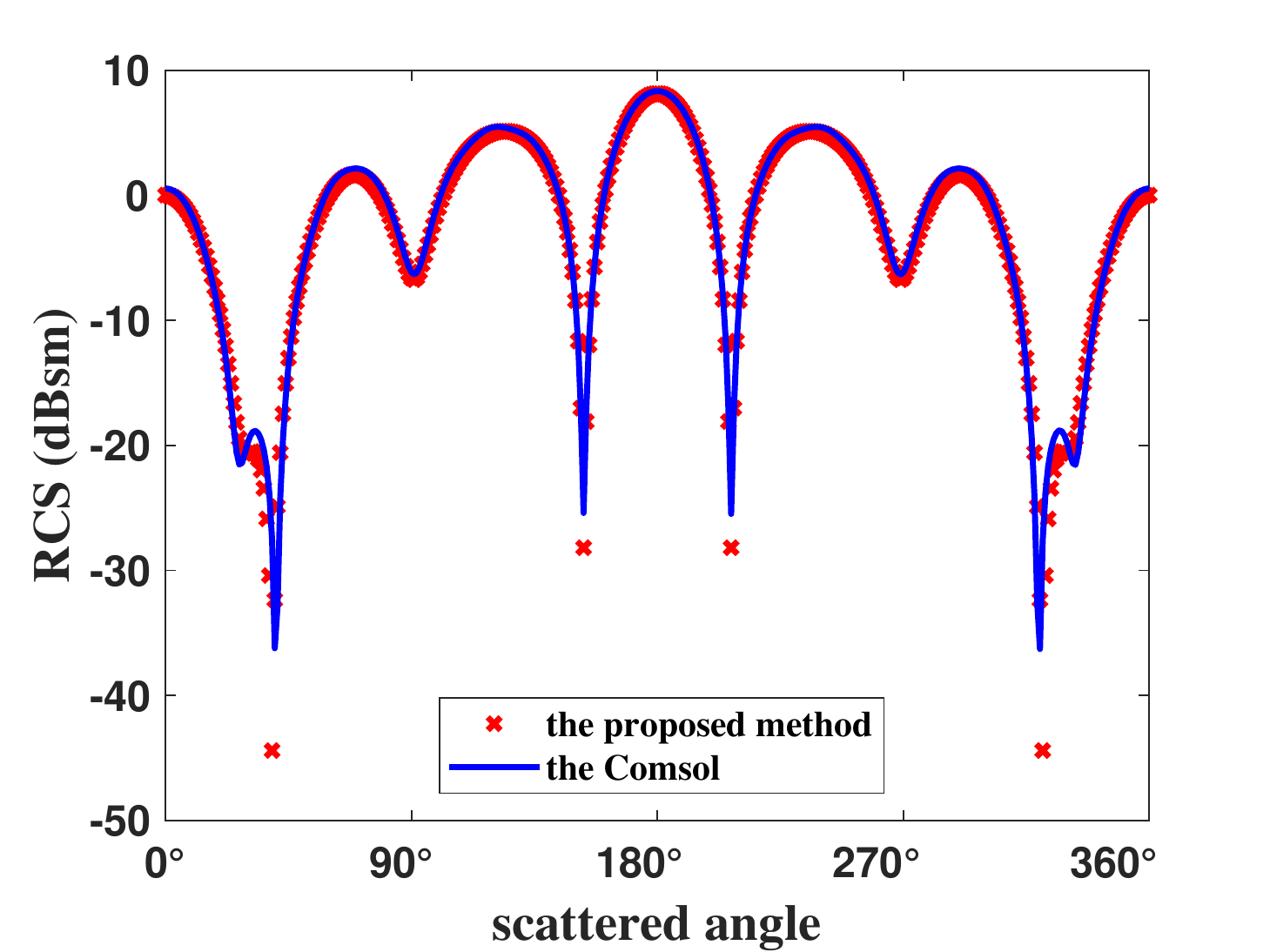}}
		\centerline{(b)}
	\end{minipage}
	\caption{(a) relative error between electric field obtained from the proposed method and Comsol and (b) RCS obtained from the proposed method and the Comsol.}
\label{fig_3}
\end{figure}

\section{Conclusion}
In this paper, we proposed a surface integral formulation for scattering modeling by 2D penetrable objects. The proposed method only requires a single electric current source derived from the surface equivalence theorem. We related the surface electric current and surface electric field through the differential surface admittance operator. However, we could obtain its dual magnetic single source formulation with a similar manner. Then, with combining the CFIE, we could solve the scattering problems by arbitrarily shaped objects and the fields also could be reconstructed from the computed surface electric fields at any location. Numerical results show that the proposed method is robust, accurate to handle smoothing and non-smoothing objects with the low to high permittivity in a wide frequency range.



\end{document}